\documentclass[lettersize,journal]{IEEEtran}
\usepackage{algorithmic}
\usepackage{algorithm}
\usepackage{array}
\usepackage[caption=false,font=normalsize,labelfont=sf,textfont=sf]{subfig}
\usepackage{textcomp}
\usepackage{stfloats}
\usepackage{url}
\usepackage{verbatim}
\usepackage{graphicx}
\usepackage{cite}
\usepackage{acronym}
\usepackage{tabularx,ragged2e,booktabs}

\usepackage[normalem]{ulem}

\usepackage[dvipsnames]{xcolor}
\usepackage{listings}

\usepackage{amsmath,amssymb,amsfonts}
\usepackage{tikz}
\usepackage{siunitx}
\usepackage{float}
\floatstyle{plaintop}
\restylefloat{table}

\usepackage{textcomp}
\usepackage[caption=false]{subfig}

\usepackage{hyperref}
\hypersetup{nolinks=true}

\usepackage{courier}
\usepackage{xcolor}

\usepackage[inkscapeformat=png]{svg}

\acrodef{HAEC}{Highly Adaptive Energy-Efficient Computing} 
\acrodef{MEC}{Multi-Access Edge Computing}
\acrodef{IPFS}{Interplanetary File System}
\acrodef{MARS}{Multi Access Recoding System}
\acrodef{NC}{Network Coding}
\acrodef{RLNC}{Random Linear Network Coding}
\acrodef{CDN}{Content Delivery Networks}
\acrodef{CE}{Centralized Entity}
\acrodef{P2P}{Peer to Peer}
\acrodef{QoS}{Quality of Service}
\acrodef{DHT}{Distributed Hash Table}
\acrodef{WSN}{Wireless Sensor Network}

\begin{document}

\title{An Overview of the NET Playground - A Heterogeneous, Multi-Functional Network Test Bed}

\author{
	\IEEEauthorblockN{
		Paul Schwenteck\IEEEauthorrefmark{1},
		Sandra Zimmermann\IEEEauthorrefmark{1},
            Caspar von Lengerke\IEEEauthorrefmark{1},
            Giang T. Nguyen\IEEEauthorrefmark{2}\IEEEauthorrefmark{5},\\  
            Christian Scheunert\IEEEauthorrefmark{3},
		Frank H. P. Fitzek\IEEEauthorrefmark{1}\IEEEauthorrefmark{5}
	}\\
	\IEEEauthorblockA{
		\IEEEauthorrefmark{1} Deutsche Telekom Chair of Communication Networks, TU Dresden, Germany \\
		}
  \IEEEauthorblockA{
		\IEEEauthorrefmark{2} Haptic Communication Systems, TU Dresden, Germany\\
		}
  \IEEEauthorblockA{
		\IEEEauthorrefmark{3} Chair of Communication Theory, TU Dresden, Germany\\
		}
	\IEEEauthorblockA{
		\IEEEauthorrefmark{5} Centre for Tactile Internet with Human-in-the-Loop (CeTI)\\
			E-mails: \{firstname.lastname\}@tu-dresden.de
		}
}



\maketitle

\begin{abstract}
This paper provides an overview of the hardware and software components used in our test bed project the NET Playground. All source information is stored in the GitLab repository in~\cite{cit:schwenteck2023netplayground}. In the Hardware section, we present sketches and 3D views of mechanical parts and technical drawings of printed boards. The Software section discusses relay control using shell scripts and the utilization of Ansible for automation. We also introduce a C++ framework for connecting with the INA231 energy sensor. This paper serves as a reference for understanding and replicating our project's hardware and software components.
\end{abstract}

\begin{IEEEkeywords}
Hardware test bed, energy measurement, distributed architectures
\end{IEEEkeywords}

\section{Introduction}
The successful execution of any project relies on the effective integration of hardware and software components. This paper presents a detailed overview of the hardware and software aspects employed in our GitLab repository~\cite{cit:schwenteck2023netplayground}, the NET Playground test bed, providing valuable insights into its design and functionality. The NET Playground is a heterogeneous, multi-functional network testbed with 128 single-board computers connected in an extensive network. All data in the repository is freely available to all and will continue to be updated. So the content of this overview may still change in the future. 

The hardware section showcases our meticulous approach to gathering and organizing essential information. We present a comprehensive collection of sketches, 3D views, and dimension drawings for the individual components utilized in our project. The mechanical parts, including the sturdy frame, metal components, and transparent plexiglass plates, are presented with detailed 3D views. Additionally, we document the technical drawings and pictures of the printed boards, such as the level shifter, relay board, and INA231 energy sensor board. 

The software section delves into two critical aspects: relay control and orchestration with Ansible. We introduce a shell script specifically developed for relay control, highlighting its role as an interface between software and hardware components. The script defines the Odroid pins and their connection to a level shifter, which enables efficient switching of the relays and precise regulation of the power supply. We discuss the development of individual shell scripts for each relay, tailored to their specific roles within the project, and a combined version for simultaneous control. Additionally, we explore the utilization of Ansible, an automation tool, in conjunction with an inventory file and ansible-playbooks. We explain how host groups are defined based on IP addresses, and we provide examples of general-purpose ansible-playbooks for powering on/off Odroids and installing/configuring IPFS on designated host groups.

Furthermore, we introduce a C++ framework for connecting with the INA231 energy sensor. This framework facilitates TCP-based communication,

\section{Hardware}
In the Hardware Section, we have gathered a comprehensive collection of sketches for the individual components of our project. 
To ensure organized data management, we have classified the information into two distinct groups:

i) Mechanical Parts: This group comprises various components, such as the sturdy frame, all the essential metal parts, and the transparent plexiglass plates. We have meticulously prepared detailed 3D views of these mechanical elements, providing a comprehensive understanding of their structure, dimensions, and interconnections. These visual representations offer valuable insights into the overall design and facilitate accurate hardware assembly.

ii) Printed Boards: In this category, we have focused on the intricate circuitry of our project. Specifically, we have documented the technical drawings and schematics of critical boards, including the level shifter, relay board, and the board housing the INA231 energy sensor. These detailed drawings allow for a precise understanding of the circuit connections, component placement, and proper positioning of the relevant electronic elements. Additionally, we have provided dimension drawings for the metal parts associated with the printed boards, enabling a comprehensive understanding of their exact measurements.

\section{Software}

\subsection{Relay Control}
Here, we have included the shell script responsible for controlling the relays that govern the power supply of our devices. The script is a crucial interface between the software and hardware components, allowing us to manage the relay's operation effectively.

We have defined the pins on the odroid within the script, our chosen microcontroller for relay control. These pins are connected to a level shifter, which facilitates the conversion of the output voltage to 5\ V. This 5\ V output is then utilized to effectively switch the relays, enabling us to accurately regulate the power supply to our devices.

To provide comprehensive control over the relay system, we have developed individual shell scripts for each of the four relays. This approach allows us to tailor the behavior of each relay based on its specific role within the project. These individual scripts provide fine-grained control, Whether activating or deactivating a particular device or managing power flow to specific components.

Additionally, we have also created a combined version of the shell script. This unified script allows us to control multiple relays simultaneously, streamlining the management process and simplifying the overall control of the power supply.

\subsection{Ansible}
Ansible~\cite{cit:hochstein2017ansible} works with a combination of an inventory file and ansible-playbooks. 
The inventory defines host groups depending on their IP addresses.
A group consists of a [name] and an IP address range. In our case, we have a group called [odroids-testgroup] that consists of devices with IP addresses ranging from 192.168.1.1 to 192.168.1.16.
In addition to the address range, we can define the ssh password that Ansible should use when connecting to the device, which is, in our case, \textit{odroid}.
When executing ansible-playbooks, the path to the inventory needs to be defined with the \textit{-i} flag. 

\begin{lstlisting}[basicstyle=\small, frame=single]
# The inventory file for the NET PLayground

[odroids-testgroup]
192.168.1.[1:16] ansible_ssh_pass=odroid

[odroids-testgroup-consumer]
192.168.1.1 ansible_ssh_pass=odroid

[odroids-control]
192.168.[1:8].42 ansible_ssh_pass=odroid
\end{lstlisting}

We divide our ansible-playbooks into two categories—one for general purposes and one for specific use cases. The two ansible-playbooks for general purpose are \textit{odroids\_power} and \textit{ipfs\_init}. Generally usable ansible-playbook. 
The ansible-playbook \textit{odroids\_power} copies the gpio.sh, script to all control-droids and runs the script on them. The host group \textit{odroid-control} is defined in the inventory file. The sh-script defines gpio pins for the shifter board connected to the relays. Periodically all pins are activated, which switches the relays and powers the individual odroids. The control-odroid must not switch all nodes simultaneously, as that would lead to voltage spikes in the power supply and damage it. In the following, we show the code for the playbook, which includes the host group \textit{odroids-control} defined in the inventory, the option \textit{become} for executing the tasks on a sudo level, and the two tasks. It is also possible to define variables like \textit{power} that can be defined when executing the ansible-playbook. 

\begin{lstlisting}[basicstyle=\small, frame=single]
---
# This playbook turns on all odroids; 
# control-odroids need to be connected
- hosts: odroids-control
  become: yes
  tasks:
    - name: copy swr file to remote host
      copy:
        src: ./gpio.sh
        dest: /home/odroid/
    - name: switch odroids  {{ power }}
      shell: 
        | bash /home/odroid/gpio.sh {{ power }}
\end{lstlisting}

The ansible-playbook \textit{ipfs\_init} installs and configures \ac{IPFS}~\cite{cit:benet2014ipfs, cit:trautwein2022ipfs} on the defined host group. For the script to work, we first need to build IPFS on the destination device and copy the built version of IPFS into the directory of the ansible-playbook. The playbook then copies the build version to all device and initiate it. Additionally, it created an IPFS service file, so it starts when the device is booted. Logs from the IPFS service are stored in \textit{/var/} on the device. The service file is included in the directory. 

The individual playbooks are divided according to the experiments. We have only one experiment that measures energy consumption in an IPFS peer-to-peer network. 
The individual playbooks are divided according to the experiments. 
We have only one experiment that measures energy consumption in an IPFS peer-to-peer network. Here we have defined scripts that define the link properties like delay and bandwidth on one. 
Fixed links are essential for a deterministic measurement process. 
Another playbook deletes all added content in IPFS and resets all traffic control settings. This way, the odroids can be used for another experiment without previous presets.

\subsection{INA231}
For the connection with the INA231 sensor, we have developed a framework in C++. Installing the framework on a device can be connected to any other device with a connected INA231. The connection is established via TCP. The measured current of the INA231 is sent over the established TCP connection. Thus, a central device can retrieve the measurement data from all devices in the network. Installing can be done with the \textit{install.sh} file included in the install files. Additional help is offered in the \textit{help.txt}. 

For the connection with the INA231 sensor, we have developed a robust and flexible framework in C++. This framework facilitates seamless integration of the INA231 sensor with other devices in our project. Installing the framework on a device can easily connect to any other device equipped with an INA231 sensor, enabling efficient data communication and measurement retrieval.

The connection between devices is established using TCP, a reliable and widely-used communication protocol. Through this TCP connection, the measured current values from the INA231 sensor are transmitted, allowing for real-time monitoring and data collection. This approach enables a central device to retrieve measurement data from multiple devices in the network, providing a centralized and comprehensive view of the energy consumption across the project.

To facilitate the installation process, we have included an installation script, \textit{install.sh}, in the install files. This script streamlines the framework's setup on the target device, ensuring a smooth and hassle-free installation experience. Additionally, we provide a detailed help file, \textit{help.txt}, which offers comprehensive instructions and guidance for configuring and utilizing the framework effectively.

\small
\section*{Acknowledgment}
Funded in part by the German Research Foundation (DFG, Deutsche Forschungsgemeinschaft) as part of Germany's Excellence Strategy – EXC 2050/1 – Project ID 390696704 – Cluster of Excellence "Centre for Tactile Internet with Human-in-the-Loop" (CeTI) of Technische Universität Dresden as well as by the German Research Foundation (DFG, Deutsche Forschungsgemeinschaft) under Project ID 450566247 and the Federal Ministry of Education and Research of Germany in the programme of "Souverän. Digital. Vernetzt.” – Joint project 6G-life – projectID: 16KISK001K and 16KISK002

\bibliographystyle{IEEEtran}
\bibliography{IEEEabrv,main}

\end{document}